\documentclass[%
 aip,
 jmp,%
 amsmath,amssymb,
preprint,
]{revtex4-1}

\usepackage{graphicx}
\usepackage{dcolumn}
\usepackage{bm}
\usepackage{verbatim}
\usepackage{float}

\begin{document}

\preprint{AIP/123-QED}

\title[B Appelbe et al]{Magnetic Field Transport in Propagating Thermonuclear Burn}

\author{B. Appelbe}
\email{bappelbe@ic.ac.uk}
\affiliation{%
 The Centre for Inertial Fusion Studies,  The Blackett Laboratory, Imperial College, London, SW7 2AZ, United Kingdom
}%

\author{A. L. Velikovich}
\affiliation{%
Plasma Physics Division, Naval Research Laboratory, Washington, DC 20375, USA
 }%

\author{M. Sherlock}
\affiliation{%
Lawrence Livermore National Laboratory, P.O. Box 808, Livermore, California 94551-0808, USA
}%

\author{C. Walsh}
\affiliation{%
Lawrence Livermore National Laboratory, P.O. Box 808, Livermore, California 94551-0808, USA
}%

\author{A. Crilly}
\affiliation{%
 The Centre for Inertial Fusion Studies,  The Blackett Laboratory, Imperial College, London, SW7 2AZ, United Kingdom
}%

\author{S. O Neill}
\affiliation{%
 The Centre for Inertial Fusion Studies,  The Blackett Laboratory, Imperial College, London, SW7 2AZ, United Kingdom
}%

\author{J. Chittenden}
\affiliation{%
 The Centre for Inertial Fusion Studies,  The Blackett Laboratory, Imperial College, London, SW7 2AZ, United Kingdom
}%

\date{\today}

\begin{abstract}
High energy gain in inertial fusion schemes requires the propagation of a thermonuclear burn wave from hot to cold fuel. We consider the problem of burn propagation when a magnetic field is orthogonal to the burn wave. Using an extended-MHD model with a magnetized $\alpha$ energy transport equation we find that the magnetic field can reduce the rate of burn propagation by suppressing electron thermal conduction and $\alpha$ particle flux. Magnetic field transport during burn propagation is subject to competing effects: field can be advected from cold to hot regions by ablation of cold fuel, while the Nernst and $\alpha$ particle flux effects transport field from hot to cold fuel. These effects, combined with the temperature increase due to burn, can cause the electron Hall parameter to grow rapidly at the burn front. This results in the formation of a self-insulating layer between hot and cold fuel that reduces electron thermal conductivity and $\alpha$ transport, increases the temperature gradient and reduces the rate of burn propagation.
\end{abstract}

\maketitle

\section{Introduction}\label{sec:1}
The ultimate goal of inertial fusion energy is to produce high energy gain, in which the energy liberated in thermonuclear reactions is orders of magnitude larger than the energy of the driver used to compress the fusion fuel. High energy gain can be achieved by using the driver energy to ignite a small fraction of the fuel, the hotspot, which then drives a propagating thermonuclear burn wave into a surrounding layer of dense, cold fuel.\cite{Lindl_1995,Slutz_PRL2012} The burn wave consists of significant energy transport due to $\alpha$ particles, electron thermal conduction and radiative processes.\cite{Tong_NF2019}

Magnetic fields are a central feature of a broad class of inertial fusion schemes referred to as Magneto-Inertial Fusion (MIF)\cite{Wurden_MIF2016}. In MIF, a magnetic field is applied to the fuel during the compression phase. This magnetic field provides \textit{magnetothermoinsulation}\cite{Lindemuth_1983} during the compression phase by reducing electron thermal conduction losses from hot plasma. This reduces the implosion velocity required to reach the ignition temperature. The magnetic field also confines $\alpha$ particles within the fuel during thermonuclear burn but with a Larmor radius that is significantly larger than that of the thermal electrons.

A number of MIF schemes under current investigation, such as MagLIF at Sandia National Laboratories\cite{Slutz_POP2010,Gomez_PRL2020}, aim to achieve volumetric thermonuclear burn. Volumetric burn schemes are designed to heat all of the fuel to fusion temperatures during the compression phase. If the fuel ignites then a net energy gain can be achieved without the need for thermonuclear burn propagation (even though the theoretical maximum energy gain is smaller for volumetric burn compared with propagating burn). For volumetric burn schemes it is desirable to have a magnetic field which maximizes the magnetothermoinsulation, as long as the magnetic pressure remains small relative to thermal pressure.

In the case of MIF schemes involving propagating burn, such as high gain MagLIF\cite{Slutz_PRL2012} and magnetized indirect-drive Inertial Confinement Fusion(ICF)\cite{Perkins_POP2013,Perkins_POP2017}, the magnetothermoinsulation effect is required during the compression phase to aid the formation of an igniting hotspot. However, the role of the magnetic field during the propagating burn phase is less clear. The suppression of thermal conduction and $\alpha$ particle transport reduces the rate of burn propagation into the cold fuel.\cite{Slutz_PRL2012,Jones_NF1986,Velikovich_ICOPS2012} This can limit the energy gain achieved before the target disassembles. While some magnetic confinement of $\alpha$ particles during burn propagation could be desirable, to allow for lower areal densities, it is clear that a magnetic field which is too large could be detrimental to the achievable yield. This question of how a magnetic field affects thermonuclear burn propagation motivates the present work. In addition to propagating burn MIF schemes, this question may also be relevant to conventional ICF since simulations have shown that the Biermann battery mechanism\cite{Biermann_1950} can generate magnetic fields at the surface of the hotspot in perturbed ICF implosions.\cite{Walsh_PRL2017}

In this work we study magnetic field dynamics in a propagating burn wave and the effect of the magnetic field on the evolution of burn propagation. The principal novelty of the work is our detailed treatment of the magnetic field transport. We utilize an MHD model (described in detail in section \ref{sec:2}) in which the induction equation includes advection, resistive diffusion, electron temperature gradients (the Nernst effect) and the flux of $\alpha$ particles. This last term is typically not included in simulations of MIF schemes. We find that the various terms in the induction equation result in competing dynamics of the magnetic field at a propagating burn front: field is advected from cold to hot regions as the burn wave causes ablation of cold fuel whilst temperature gradients and $\alpha$ particle fluxes transport field in the opposite direction. These dynamics have a significant effect on the shape and rate of propagation of the burn wave.

The magnetic field profile at the onset of burn in MIF is determined by the implosion phase and there has been a number of studies showing the complex field transport occurring in that phase.\cite{Velikovich_POP2015, Velikovich_POP2019, GarciaRubio_PRE2018,GarciaRubio_POP2017} For simplicity, we do not study the implosion phase in the current work. Instead we take as our starting point a region of hot fuel that is undergoing significant $\alpha$ heating adjacent to a cold fuel layer. In order to elucidate the field dynamics during burn, we focus on three simple cases for the initial magnetic field profile: a spatially uniform field, a spatially uniform electron Hall parameter and a magnetic field that is non-zero only in a small region at the burn front. Results for each of these cases are discussed in section \ref{sec:3}.

This is not an exhaustive search of magnetic field strengths and profiles but the cases considered show that there is a complex interplay between the magnetic field dynamics and the evolution of the propagating burn wave. The suppression of thermal conduction and $\alpha$ transport by the magnetic field depends on both the magnitude of the magnetic field and the collisionality of the plasma, where the relevant metric is the Hall parameter. As thermonuclear burn heats the plasma, collisionality decreases rapidly. Combining this with the magnetic field transport effects leads to enhanced suppression of energy transport from hot to cold fuel. In our case studies we identify scenarios in which a self-insulating magnetized layer can form between hot and cold fuel due to feedback between decreasing collisionality in the burn wave and the magnetic field transport. This self-insulating layer significantly reduces thermal conductivity and $\alpha$ transport across it and leads to the development of large temperature gradients. In section \ref{sec:4} we formulate an equation for the growth rate of the Hall parameter in order to better understand the conditions leading to formation of the self-insulating layer. This equation illustrates, inter alia, the dependence of the growth rate on the Hall parameter value, $\alpha$ heating, and the temperature and $\alpha$ energy density profiles.

Finally, we present some conclusions in section \ref{sec:5}. While recent theoretical\cite{Velikovich_POP2019} and experimental\cite{Gomez_PRL2020} work has demonstrated the importance of magnetic field transport during the implosion phase, our results show that field transport effects during the burn phase are also of interest.

\section{Model Outline}\label{sec:2}
Our model is a one-dimensional, planar, classical DT plasma consisting of semi-infinite regions of cold and hot fuel, separated by a smooth transition region with an orthogonal magnetic field. This idealized geometry is not intended to model a specific MIF scheme, but instead allows us to study how thermonuclear burn propagates from hot to cold fuel. The system is governed by an equation of continuity, the isobaric condition, an induction equation and energy equations for the fuel and the $\alpha$ particles as follows:
\begin{eqnarray}
\frac{\partial n}{\partial \hat{t}}+\frac{\partial}{\partial \hat{x}}\left(n\hat{u}\right)&=&0,\label{e:1}\\
\frac{\partial}{\partial \hat{x}}\left(2nT+\frac{B^{2}}{2\mu_{0}}\right)&=&0,\label{e:2}\\
\frac{\partial B}{\partial \hat{t}}+\underbrace{\frac{\partial}{\partial \hat{x}}\left(\hat{u}B\right)}_\text{advect.}&=&\frac{\partial}{\partial \hat{x}}\left[\underbrace{\hat{\alpha}\frac{\partial B}{\partial \hat{x}}}_\text{res. diff.}+\left(\underbrace{\hat{\beta}\frac{\partial \ln T}{\partial \hat{x}}}_\text{Nernst}+\underbrace{\hat{\gamma}\frac{\partial \ln \mathcal{E}_{\alpha}}{\partial \hat{x}}}_\text{$\alpha$-e coll.}\right)B\right],\label{e:3}\\
3n\frac{D T}{D \hat{t}}+2nT\frac{\partial \hat{u}}{\partial \hat{x}}&=& \frac{\partial}{\partial \hat{x}}\left[\underbrace{\hat{\kappa}\frac{\partial T}{\partial \hat{x}}}_\text{cond.}+\underbrace{\hat{\beta}\frac{B}{\mu_{0}}\frac{\partial B}{\partial \hat{x}}}_\text{Ettings.}\right]+\underbrace{\hat{q}_{\alpha}}_\text{$\alpha$ heat.}-\underbrace{\hat{P}}_\text{brem.},\label{e:4}\\
\frac{D \mathcal{E}_{\alpha}}{D \hat{t}}+\frac{5}{3}\mathcal{E}_{\alpha}\frac{\partial \hat{u}}{\partial \hat{x}} &=& \underbrace{\frac{\partial}{\partial \hat{x}}\left(\hat{\delta}_\mathcal{E}\frac{\partial \mathcal{E}_{\alpha}}{\partial \hat{x}}\right)}_\text{$\alpha$ energy diff.}-\hat{q}_{\alpha}+\underbrace{\hat{Q}}_\text{reac.},\label{e:5}
\end{eqnarray}
where $\hat{t}$, $\hat{x}$ and $\hat{u}$ represent dimensionless time, position and fluid velocity of the fuel and
\begin{equation}
\frac{D }{D \hat{t}} = \frac{\partial }{\partial \hat{t}}+\hat{u}\cdot\frac{\partial }{\partial \hat{x}},\nonumber
\end{equation} 
is the convective derivative. The normalizing constants are $L_{T}$, representing the width of the transition region between hot and cold fuel and $\mathcal{T}_{h}$, the stopping time for a $3.45\,MeV$ $\alpha$ particle in the initial hot fuel. Other variables include fuel density, $n$, magnetic field, $B$, fuel temperature, $T$, and $\mathcal{E}_{\alpha}$ is the energy density of $\alpha$ particles. SI units are used for quantities that have not been made dimensionless.

The induction, \eqref{e:3}, and fuel energy, \eqref{e:4}, equations contain classical magnetized transport coefficients for resistivity, thermoelectricity and thermal conductivity\cite{Braginskii_1965}
\begin{eqnarray}
\hat{\alpha} &=& \frac{\mathcal{T}_{H}}{L_{T}^{2}}\frac{m_{e}}{\mu_{0}e^{2} n\tau_{ei}}\alpha_{\bot}^{c},\qquad\hat{\beta} = \frac{\mathcal{T}_{H}}{L_{T}^{2}}\frac{\tau_{ei}T}{m_{e}}\frac{\beta_{\wedge}}{\chi_{e}},\nonumber\\
\hat{\kappa} &=& \frac{\mathcal{T}_{H}}{L_{T}^{2}}\frac{n T \tau_{ei}}{m_{e}}\left(\kappa_{\bot e}^{c}+ \sqrt{\frac{2m_{e}}{m_{i}}}\kappa_{\bot i}^{c}\right),\nonumber
\end{eqnarray}
The coefficients $\alpha_{\bot}^{c}$, $\beta_{\wedge}$, $\kappa_{\bot e}^{c}$ and $\kappa_{\bot i}^{c}$ are functions of electron Hall parameter, $\chi_{e} = eB\tau_{ei}/m_{e}$, where
$\tau_{ei} = 6\varepsilon_{0}^{2}\sqrt{2\pi^{3}m_{e}}T^{\frac{3}{2}}/\left(e^{4}\ln\Lambda_{ei} n\right)$. We use the tabulated values of Epperlein and Haines\cite{Epperlein_1986} for these coefficients. The chosen geometry of our problem means that the Hall term\cite{Farmer_2019} and Biermann battery term\cite{Biermann_1950} do not need to be considered in the induction equation.

The final term on the rhs of \eqref{e:3} represents the collisionally-induced current arising from the interaction of the $\alpha$ particle flux with the thermal electrons. Similar fast ions effects have previously been studied for collisionless plasmas in current drive in tokamaks\cite{Fisch_RevModPhys1987}, cosmic rays in astrophysical plasmas\cite{Bai_ApJ2015} and relativistic particles in laser-plasma interactions.\cite{Sherlock_PRL2010} Here we follow the procedure of Appelbe et al\cite{Appelbe_POP2019} to include this effect in a collisional, magnetized plasma. 

An analogy may be drawn between the Nernst and the $\alpha$-e collisional terms in \eqref{e:3}. The Nernst effect is a result of the effect of a magnetic field on the thermal force.\cite{Braginskii_1965} Electrons being driven by a temperature gradient and experiencing friction due to the background ions are deflected by a magnetic field that is orthogonal to the temperature gradient, generating an electron current orthogonal to both. In the case of the $\alpha$-e collisional term the electron current is instead driven by a flux of $\alpha$ particles. Both the electron temperature gradient and the $\alpha$ flux are themselves dependent on the magnetic field. The interplay between temperature gradient, magnetic field and driven electron current has been well studied\cite{Velikovich_POP2019} and is often included in simulations. However, the equivalent interplay between the magnetic field, the $\alpha$ flux and the driven electron current has not received such attention.

In our geometry the $\alpha$-e collisional effect can be expressed as
\begin{eqnarray}
\hat{\gamma}&=&\mathcal{E}_{\alpha}\left(\frac{\partial \mathcal{E}_{\alpha}}{\partial \hat{x}}\right)^{-1}\frac{Z_{\alpha}}{n}\left(\gamma_{\bot}F_{\bot}+\gamma_{\wedge}F_{\wedge}\right),\label{e:7}\\
\gamma_{\bot} &=& -j_{\bot}^{\nu}\left(1+\frac{\alpha_{\wedge}^{c}}{\chi_{e}}\right)+j_{\wedge}^{\nu}\frac{\alpha_{\bot}^{c}}{\chi_{e}},\label{e:7b}\\
\gamma_{\wedge} &=& j_{\wedge}^{\nu}\left(1+\frac{\alpha_{\wedge}^{c}}{\chi_{e}}\right)+j_{\bot}^{\nu}\frac{\alpha_{\bot}^{c}}{\chi_{e}},\label{e:7c}
\end{eqnarray}
where $F_{\bot}$, $F_{\wedge}$ are components of $\alpha$ particle flux in the directions of burn propagation and orthogonal to burn propagation and $B$ field, respectively, and $j_{\bot}^{\nu}$, $j_{\wedge}^{\nu}$ are functions of $\chi_{e}$.\cite{Appelbe_POP2019} 

The $\alpha$ particle fluxes can be estimated from the gradients of $\mathcal{E}_{\alpha}$ and $\alpha$ particle diffusion coefficients as follows\cite{liberman_velikovich_1984}
\begin{eqnarray}
F_{\bot,\wedge} &=& -\frac{\hat{\delta}_{p\bot,\wedge}}{\langle E_{\alpha}\rangle}\frac{\partial  \mathcal{E}_{\alpha}}{\partial \hat{x}},\label{e:8}\\
 \hat{\delta}_{p\bot} &=& \frac{\mathcal{T}_{h}}{L_{T}^{2}}\frac{E_{\alpha 0}\tau_{\alpha}}{3m_{\alpha}\left(1+\chi_{\alpha}^{2}\right)},\quad \hat{\delta}_{p\wedge} = \chi_{\alpha} \hat{\delta}_{p\bot}\nonumber
\end{eqnarray}
where $\langle E_{\alpha}\rangle \approx E_{\alpha 0}/2$ is the mean $\alpha$ particle energy.

The $\alpha$ energy transport equation, \eqref{e:5}, represents a single group diffusion approximation.\cite{liberman_velikovich_1984}  It contains terms for magnetized diffusion, slowing on electrons and a source due to DT reactions \cite{BoschHale_1992}
\begin{eqnarray}
\hat{\delta}_{\mathcal{E}} &=& \frac{\mathcal{T}_{h}}{L_{T}^{2}}\frac{E_{\alpha 0}\tau_{\alpha}}{m_{\alpha}\left(9+\chi_{\alpha}^{2}\right)},\quad \hat{\nu} = \frac{\mathcal{T}_{h}}{\tau_{\alpha}},\quad
\hat{Q} = \mathcal{T}_{h}\frac{n^{2}}{4}\langle\sigma v\rangle_{DT}E_{\alpha0},\nonumber\\
\chi_{\alpha} &=& \frac{Z_{\alpha}eB}{m_{\alpha}}\tau_{\alpha},\quad
\tau_{\alpha} = 3\sqrt{\frac{\pi^{3}}{2m_{e}}}\frac{m_{\alpha}\varepsilon_{0}^{2}}{e^{4}\ln\Lambda_{\alpha e}}\frac{T^{\frac{3}{2}}}{n},\nonumber
\end{eqnarray}
where $\chi_{\alpha}$ is the $\alpha$ Hall parameter, $\tau_{\alpha}$ is slowing time on electrons and $E_{\alpha 0}=3.45\,MeV$. Electron and $\alpha$ Hall parameters are related by $\chi_{\alpha}/\chi_{e}=Z_{\alpha}\ln\Lambda_{ei}/\left(4\ln\Lambda_{\alpha e}\right)$. Finally, $\hat{q}_{\alpha}=2\hat{\nu}\mathcal{E}_{\alpha}$ and bremsstrahlung losses\cite{NRL_form} are $\hat{P} = 1.69\times 10^{-38}n^{2}\sqrt{\frac{T}{e}}\mathcal{T}_{h}\,J\,m^{-3}$.

Returning to the induction equation, \eqref{e:3}, we note that it has been written such that $\hat{\alpha}$, $\hat{\beta}$ and $\hat{\gamma}$ can all be treated as diffusion coefficients. In the burning plasma regime the magnetic Lewis number,\cite{Velikovich_POP2015} defined as $
Lm = \left.\hat{\kappa}/\left(3n\hat{\alpha}\right)\right|_{\chi_{e}=0},$
is large, meaning that thermal diffusivity is much larger than magnetic diffusivity and so resistive diffusion plays only a minor role in $B$ field transport. Using \eqref{e:8} for $\alpha$ particle fluxes we can write \eqref{e:7} more compactly as
\begin{equation}
\hat{\gamma} = \frac{\mathcal{T}_{h}}{L_{T}^{2}}\frac{Z_{\alpha}\mathcal{E}_{\alpha}E_{\alpha 0}\tau_{\alpha}}{3m_{\alpha}n\langle E_{\alpha}\rangle}\underbrace{\left[\frac{-\gamma_{\bot}-\chi_{\alpha}\gamma_{\wedge}}{1+\chi_{\alpha}^{2}}\right]}_\text{$\Gamma_{\chi}$},\nonumber
\end{equation}
where the dependency on magnetization is now fully contained in the square brackets which we now denote as $\Gamma_{\chi}$. We note that even though \eqref{e:7b} and \eqref{e:7c} contain terms $\propto 1/\chi_{e}$, the term $\Gamma_{\chi}$ does not diverge at $\chi_{e}= 0$ since $\alpha_{\wedge}^{c},j_{\wedge}^{\nu},F_{\wedge} \rightarrow 0$ as $\chi_{e}\rightarrow 0$.

\begin{figure}[H]
\begin{center}%
\includegraphics*[width=1.0\columnwidth]{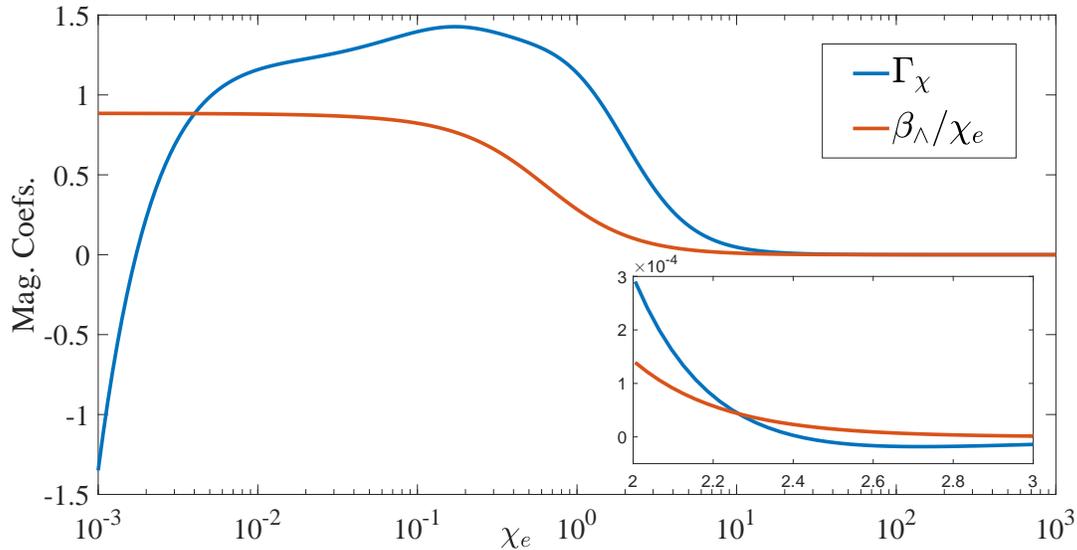}%
\vspace{-1em}
\end{center}
\caption[]{The variation of the diffusion coefficients $\hat{\beta}$ and $\hat{\gamma}$ with $\chi_{e}$, where we assume $\chi_{\alpha}=\chi_{e}/2$. Note that $\beta_{\wedge}/\chi_{e}\geq 0\,\forall\, \chi_{e}$ while $\Gamma_{\chi}<0$ for $\chi_{e}\ll 1$ and $\chi_{e}\gg 1$.  For $\chi_{e}=0$ we have $\Gamma_{\chi}\approx-1.7$. The scale of the horizontal axis of the inset plot represents $\log_{10}\chi_{e}$.} \label{f:7}
\end{figure}

The $\Gamma_{\chi}$ term is compared with the dependence of $\hat{\beta}$ (which determines the magnitude of the Nernst effect) on $\chi_{e}$ in fig. \ref{f:7}. Comparing the unmagnetized components of $\hat{\beta}$ and $\hat{\gamma}$ then gives
\begin{equation}\label{e:11}
\frac{\hat{\gamma}}{\hat{\beta}} \sim \frac{E_{\alpha 0}}{6 T}\frac{n_{\alpha}}{n}\frac{\ln\Lambda_{ei}}{\ln\Lambda_{\alpha e}}\Gamma_{\chi}\left(\frac{\chi_{e}}{\beta_{\wedge}}\right),
\end{equation}
where $n_{\alpha}$ is the number density of non-thermalised $\alpha$ particles. We can assume that the ratio of Coulomb Logarithms is of order unity and $E_{\alpha 0}/\left(6 T\right)\sim10^{2}$ for an igniting plasma. Therefore, a relative density of $\alpha$ particles of $n_{\alpha}/n\sim10^{-2}$ could be sufficient for the $\hat{\gamma}$ term to dominate.

It is also worth noting that $\hat{\beta}\geq 0$ for all values of $\chi_{e}$. This is because in a plasma the friction force exerted by background ions is lower for faster electrons. However, $\Gamma_{\chi}<0$ for very large and very small values of $\chi_{e}$. This is due to $j_{\bot}^{\nu} < 0$ at small values of $\chi_{e}$, while for large values the $-\chi_{\alpha}\gamma_{\wedge}$ term dominates.

The system of equations \eqref{e:1}-\eqref{e:5} is solved using an implicit Lagrangian model in order to accurately resolve the steep gradients that exist at the burn front.\cite{Velikovich_POP2015, Velikovich_POP2019} We use the isobaric condition, \eqref{e:2}, in place of a momentum equation for simplicity as our principal interest is the transport processes occurring in \eqref{e:3}-\eqref{e:5}. However, this limits our studies to burn propagation regimes driven by deflagration rather than detonation.

\section{Burn propagation dynamics for various $B$ profiles}\label{sec:3}
In this section we apply the model to a variety of different initial $B$ field profiles. In each case the initial temperature profile is chosen to be sigmoidal
\begin{equation}
T_{0}\left(\hat{x}\right) = T_{c}+\left(T_{h}-T_{c}\right)\left(1+\exp\left(-10\hat{x}\right)\right)^{-1},\nonumber\label{e:6}
\end{equation}
where $T_{c}$ ($T_{h}$) is the temperature of the cold (hot) fuel. The factor of $10$ ensures that $98$\% of the temperature change takes place within a region of unity length. The normalizing constant, $L_{T}$, is the burn front scale length. Its value is chosen to be some multiple of the stopping distance of a $3.45\,MeV$ $\alpha$ particle at the conditions of the midpoint of the initial temperature profile.

\subsection{Uniform $B$ field}\label{sec:3.1}
\begin{figure}[H]
\begin{center}%
\includegraphics*[width=1.0\columnwidth]{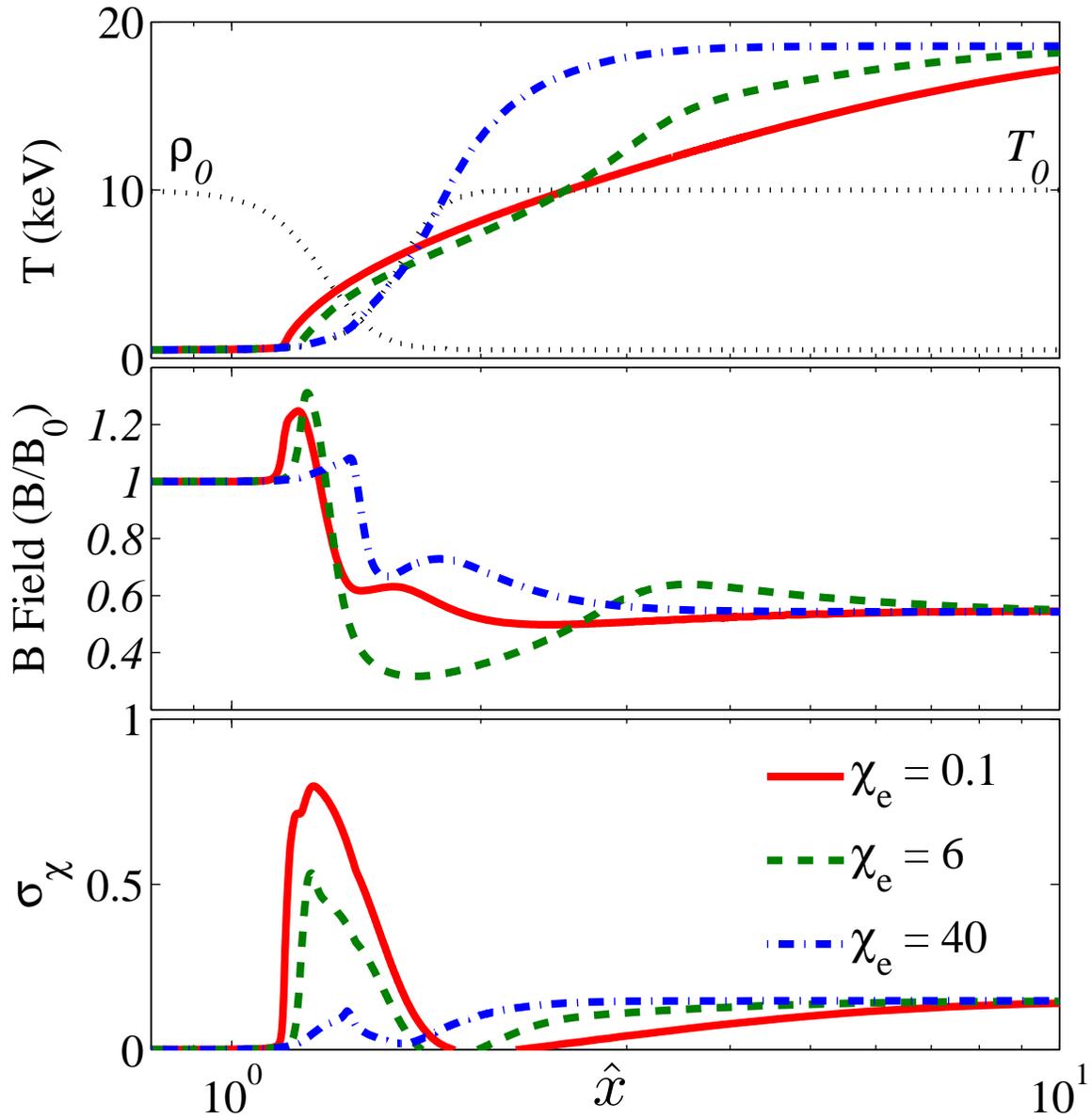}%
\vspace{-1em}
\end{center}
\caption[]{Profiles at $\hat{t}=5\mathcal{T}_{h}$ for initially uniform $B$ field with $\chi_{e} = 0.1,6,40$ in the hot fuel. Top: Temperatures. Initial temperature, $T_{0}$, and density, $\rho_{0}$, are also shown. Middle: $B$ fields where $B_{0} \approx 80, 500, 3500\,T$ for $\chi_{e} = 0.1,6,40$, respectively. Bottom: Growth rates of $\chi_{e}$. Parameter values of $T_{c} = 500\,eV$, $T_{h} = 10\,keV$, $n_{h} = 10^{31}\,m^{-3}$ and $L_{T}=0.3$ were used. Note that the isobaric condition means that the self-heating hot fuel is allowed to freely expand, resulting in $B$ decreasing uniformly in the hot fuel as $T$ increases.} \label{f:1}
\end{figure}

We begin with the case of a $B$ field whose strength is initially uniform across the burn front. Since $B$ is constant, $\chi_{e}\propto \tau_{ei}$, and the initial value of $\chi_{e}$ will vary significantly across the burn front with the largest value in the hot fuel. Figure \ref{f:1} shows results for three different initial values of $B$ in which $L_{T}$ is $0.3$ times the $\alpha$ stopping distance, corresponding to $\sim15\,\mu m$. The initial values of $\chi_{e}$ in the hot fuel were $0.1,6,40$, corresponding to $B_{0} \approx 80, 500, 3500\,T$. The corresponding values in the cold fuel were a factor of $\sim800$ smaller. We note that at $\chi_{e} =0.1,\, 6,\,40$, the electron thermal conductivity is a factor of $\sim 1,\,10^{-1}\,10^{-3}$ lower than its unmagnetized value. The maximum mean free path of thermal electrons in this system is $\lambda_{ei}\approx0.009$, ensuring validity for our MHD model.

As burn propagates the evolution of the burn front is highly sensitive to the $B$ field. The $\chi_{e}=0.1$ case (which closely resembles an unmagnetized burn front) develops a temperature profile that is far less steep than the more magnetized cases and propagates furthest into the cold fuel. However, this burn front also contains a localized region of steep temperature gradient near the cold fuel, which we refer to as the foot of the burn front. This is caused by rapid $\alpha$ heating of the cold, dense fuel. In the more magnetized cases, the burn propagates less far due to suppression by the magnetic field of thermal conduction and $\alpha$ transport from hot to cold fuel.

Transport of the $B$ field during burn is also dependent on $\chi_{e}$. The middle diagram of fig. \ref{f:1} illustrates competing transport effects. The Nernst and $\alpha$ flux effects transport $B$ field from hot to cold plasma. This is most clearly evidenced by the spikes at the foot of the burn front where temperature gradients rapidly increase. This is similar to the Nernst waves that have previously been studied in MagLIF and magnetized laser ablation.\cite{Velikovich_POP2019,Nishiguchi_PRL1984} Magnetic resistivity limits the sharpness of these spikes. Meanwhile, ablation of cold fuel by the burn front leads to $B$ field being advected from cold to hot fuel. Evidence of the ablative effect is shown in the $\chi_{e}=6$ case, where a dip in $B$ field centered at $\hat{x}=1.7$ coincides with a shallower temperature gradient. 

\begin{figure}[H]
\begin{center}%
\includegraphics*[width=1.0\columnwidth]{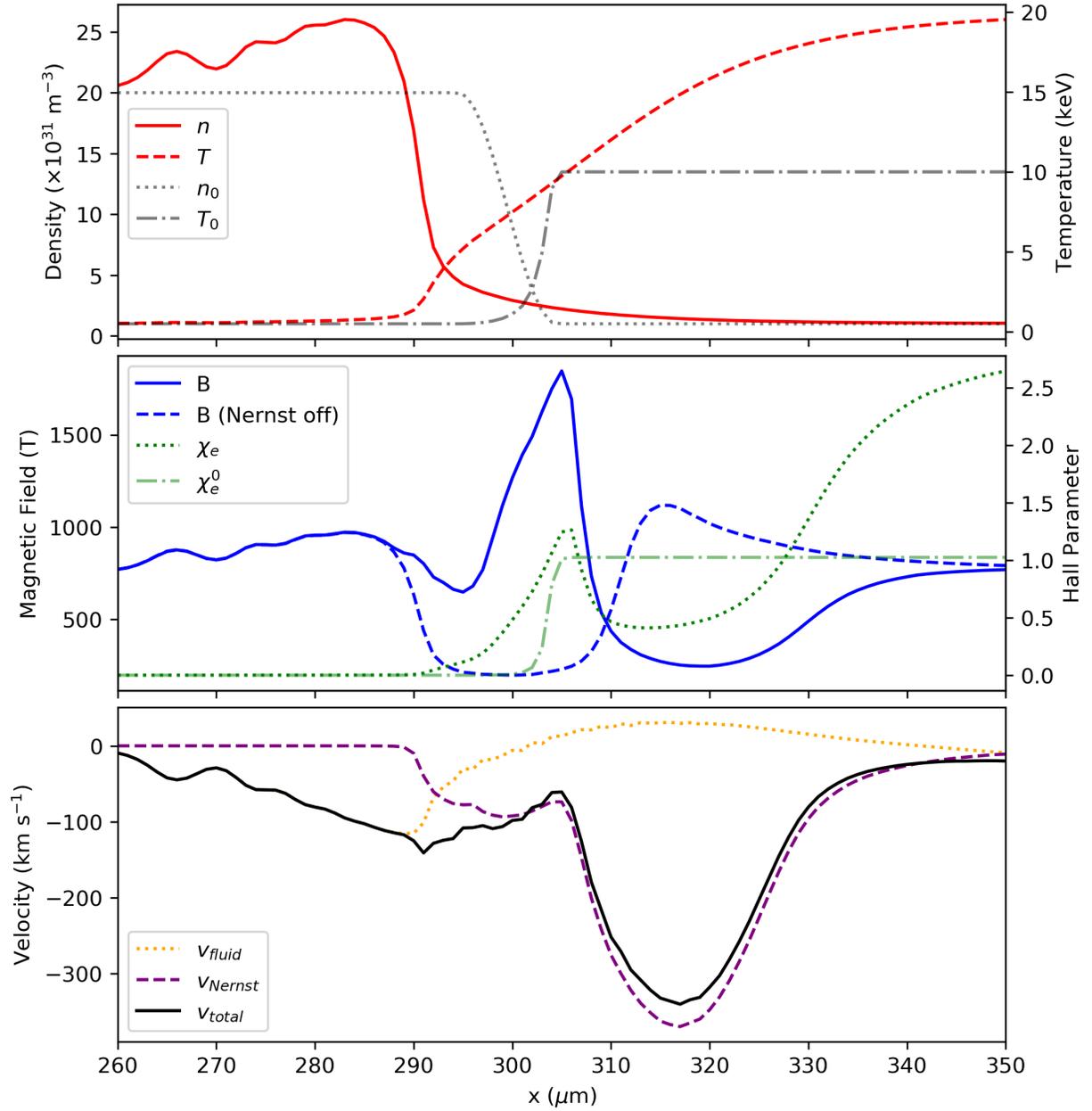}%
\vspace{-1em}
\end{center}
\caption[]{Results from an integrated simulation with an initially uniform $B$ field. Top: Initial and final profiles of temperature and density. Middle: Final profiles of $B$ field with and without the Nernst term included. Also, initial and final $\chi_{e}$ profiles. Bottom: $B$ field advection velocity ($v_{fluid}$), Nernst velocity ($v_{Nernst}$) and their sum.} \label{f:11}
\end{figure}

The dynamics at an evolving burn front can cause both compression and rarefaction of a magnetic field. Due to the isobaric condition and the increasing temperature of the plasma, advection does not cause any compression of $B$ field, only rarefaction.  However, the Nernst effect also contributes to rarefaction of the $B$ field in regions behind the compression wave. This effect has previously been observed in the compression phase of MagLIF\cite{Slutz_POP2010} and will be more clearly shown in fig. \ref{f:11}.

The bottom diagram of fig. \ref{f:1} shows profiles for $\sigma_{\chi}$, the exponential growth rate for $\chi_{e}$ at $\hat{t}=5\mathcal{T}_{h}$. The value of $\chi_{e}$ can increase due to both heating of the plasma (which increases $\tau_{ei}$) and $B$ field transport. In the cases of $\chi_{e} = 0.1$ and $6$, peak values of $\sigma_{\chi}$ occur near the foot of the burn front and are significantly larger than the $\sigma_{\chi}$ values in the hot fuel. In the $\chi_{e}=40$ case, the temperature at the foot of the burn front is increasing at a slower rate and so the value of $\sigma_{\chi}$ is approximately the same as that in the hot fuel.

\begin{figure}[H]
\begin{center}%
\includegraphics*[width=1.0\columnwidth]{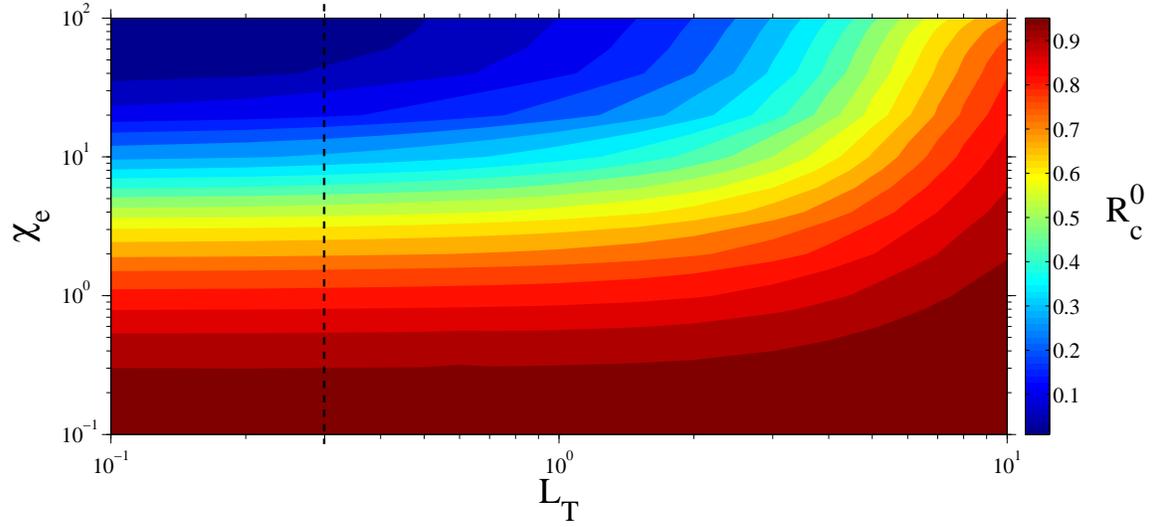}%
\vspace{-1em}
\end{center}
\caption[]{Ratio of burn rate in cold fuel to the unmagnetized burn rate at $\hat{t}=5\mathcal{T}_{h}$. Cold fuel is defined as all fuel with an initial temperature less than $\left(T_{c}+T_{h}\right)/2$. The $B$ field is initially uniform and $\chi_{e}$ denotes the initial value in the hot fuel. Values of $T_{c}$, $T_{h}$ and $n_{h}$ are as in fig. \ref{f:1}. The dashed black line indicates $L_{T} = 0.3$, the cases shown in fig. \ref{f:1}.} \label{f:2}
\end{figure}

The effects of these transport phenomena on burn propagation are illustrated in fig. \ref{f:2}. There is a significant drop in the rate of burn propagation as the initial value of $\chi_{e}$ increases. The hot fuel is insulated from the cold fuel and, since $\sigma_{\chi}$ is large at the burn front, this insulating effect increases as burn evolves. Similar behaviour is found for a wide range of initial parameters ($n_{h}\sim\left[10^{28},5\times 10^{31}\right]\,m^{-3}$ and $T_{h}>\sim 5\,keV$).

We conclude this subsection with the results of an integrated simulation, see fig. \ref{f:11}, carried out using the Chimera code\cite{Tong_NF2019} which includes radiation transport, extended-MHD and $\alpha$ transport package. However, the $\alpha-e$ collisional term has not yet been included in this model. These simulations used a similar set-up and initial conditions to those shown in fig. \ref{f:1} but with an initial value of $\chi_{e}\approx 1$ in the hot fuel.

The Nernst term causes a significant spike in the $B$ field at the burn front, with a corresponding spike in $\chi_{e}$, by compressing the $B$ field. A comparison of the $B$ field profiles with and without Nernst, shown in the middle panel, demonstrates how the Nernst also causes rarefaction of the $B$ field behind this compression wave. The bottom panel of fig. \ref{f:11} shows the velocity components of $B$ field transport with $v_{fluid}$ ($v_{Nernst}$) equivalent to $\hat{u}$ ($\hat{\beta}\partial \ln T/\partial \hat{x}$) in \eqref{e:3}. From this we can see that the spike in $B$ field corresponds to a region in which the absolute value of $v_{Nernst}$ has a local minimum. Although the isobaric condition no longer applies, the evolution of the profiles of $\rho$, $T$ and $B$ is qualitatively similar to the results obtained from the model outlined in section \ref{sec:2} and suggest that observations from this model are not unduly compromised by the isobaric condition.

\newpage

\subsection{Uniform electron Hall parameter}\label{sec:3.2}
\begin{figure}[H]
\begin{center}%
\includegraphics*[width=1.0\columnwidth]{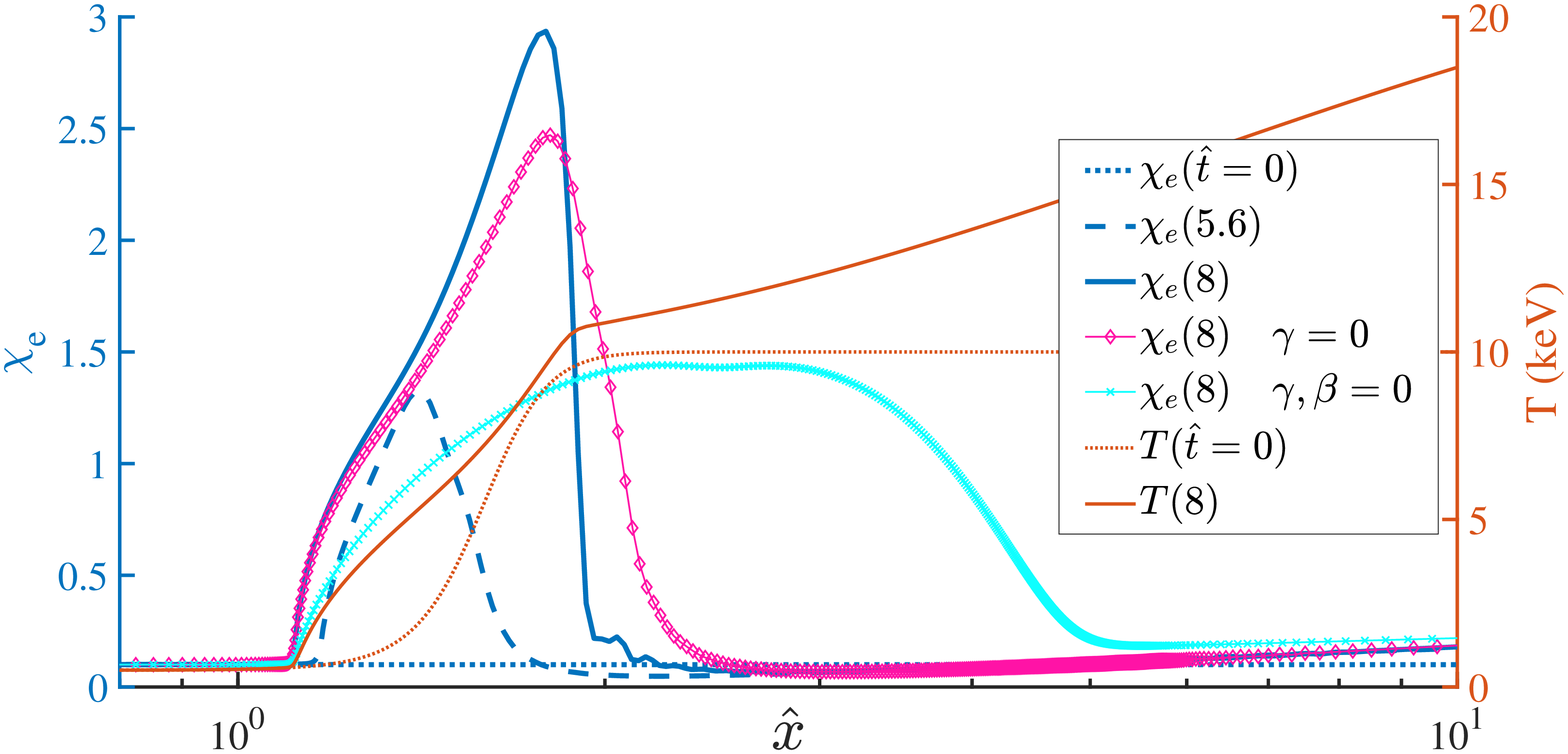}%
\hfil%
\includegraphics*[width=1.0\columnwidth]{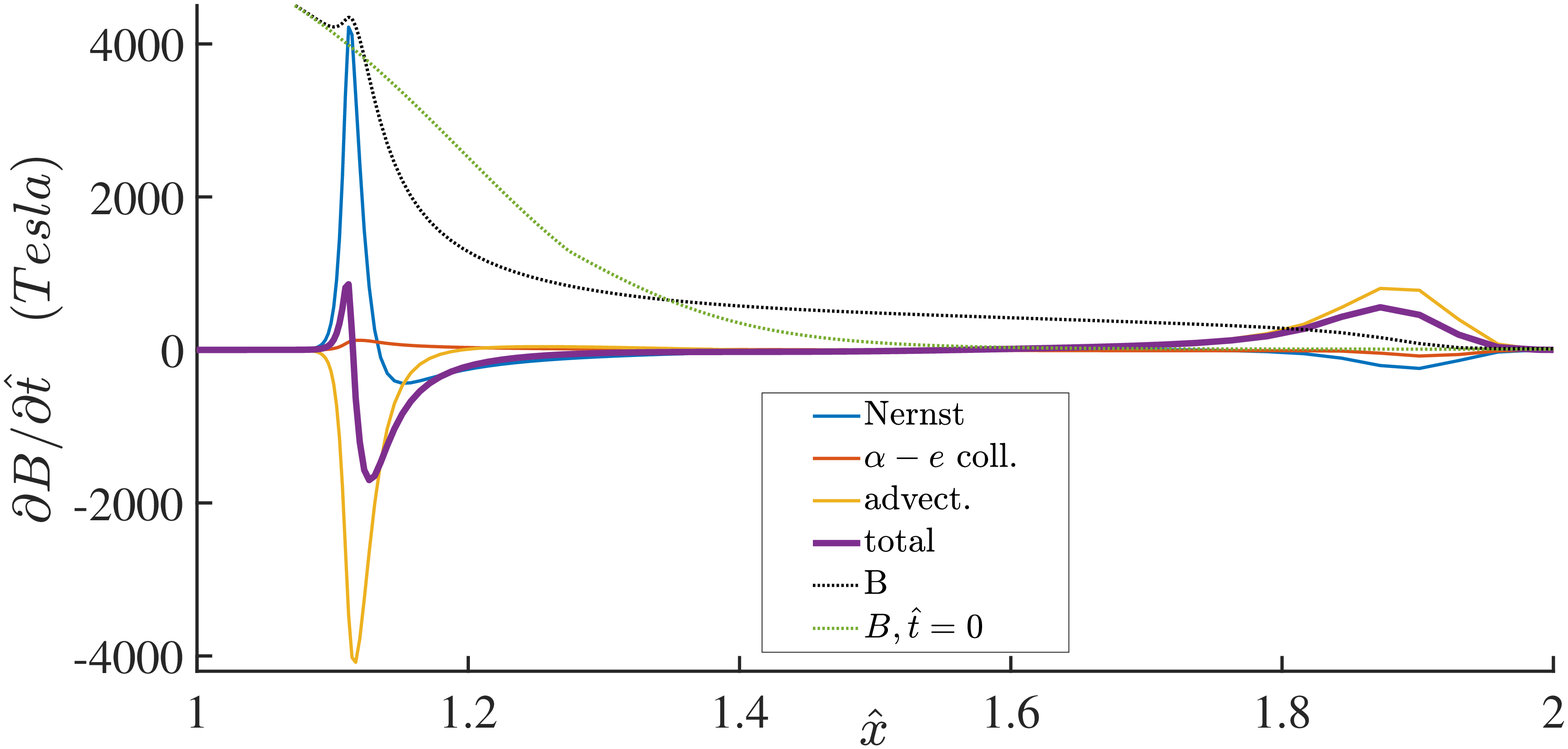}%
\vspace{-1em}
\end{center}
\caption[]{Top: Evolution of $\chi_{e}$ over a period of $8\mathcal{T}_{h}$ given an initial value $\chi_{e}=0.1$. Also shown are initial and final $T$ profiles, and the final $\chi_{e}$ profiles when the (i) $\alpha-e$ collision term (labelled $\gamma = 0$) and (ii) $\alpha-e$ and Nernst terms (labelled $\gamma, \beta = 0$) are neglected. Bottom: Components of $B$ field growth rate at the final time in the self-insulating region. Also shown are initial and final $B$ field profiles. Parameters: $T_{c} = 500\,eV$, $T_{h} = 10\,keV$, $n_{h} = 10^{30}\,m^{-3}$, $L_{T}=0.3$.} \label{f:3}
\end{figure}

We next consider the case of a burn front in which $\chi_{e}$ is initially uniform. Figure \ref{f:3} shows an example of burn propagation where initially $\chi_{e}=0.1$. To achieve uniformity of $\chi_{e}$, the initial $B$ field needs to be significantly larger in the cold fuel and so we choose a parameter value of $n_{h} = 10^{30}\,m^{-3}$ to ensure that the magnetic pressure is small compared with thermal pressure.

The top diagram of fig. \ref{f:3} illustrates how $\chi_{e}$ grows rapidly in the burn front region due to both increasing temperature and increasing magnetic field. The bottom diagram shows different terms contributing to magnetic field growth in the burn front region.

\begin{figure}[H]
\begin{center}%
\includegraphics*[width=1.0\columnwidth]{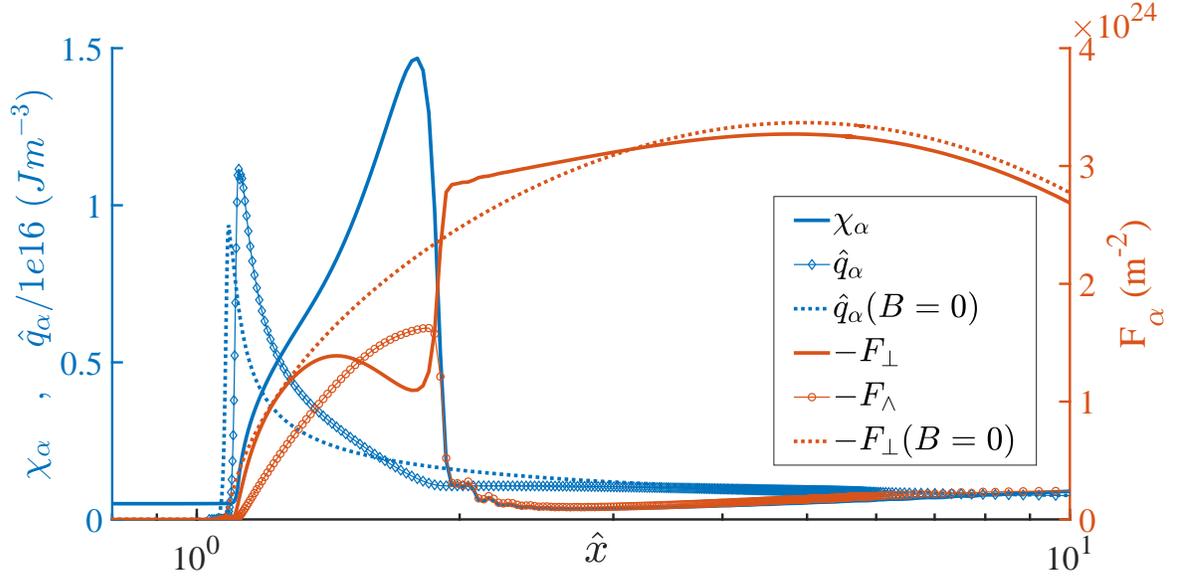}%
\vspace{-1em}
\end{center}
\caption[]{Profiles of $\chi_{\alpha}$, $\alpha$ heating and $\alpha$ flux terms after $8\mathcal{T}_{h}$. Also shown are the profiles for the case with no $B$ field. The Larmor radius for $\alpha$ particle with energy $\langle E_{\alpha}\rangle$ at maximum $\chi_{\alpha}$ is $\approx 4$ and its mean free path is $\approx 0.11$.} \label{f:4}
\end{figure}

The final $T$ profile has a distinctive shape with very steep temperature gradients developing at two locations. First, at the foot of the burn pulse where $\alpha$ heating is largest (see fig. \ref{f:4}) and secondly in the region where $\chi_{e}$ reaches a maximum. This second steep temperature gradient evolves to compensate for the drop in electron thermal conductivity caused by increasing $\chi_{e}$. This temperature profile is reminiscent of double ablation front structures\cite{Drean_POP2010,Sanz_POP2009} which occur in direct-drive ICF. In that scenario radiation energy and electron heat flux are absorbed at different locations, leading to the formation of the double ablation front. In our case the first front is formed by $\alpha$ heating while the second front is due to the suppression of energy transport.

At the location of the second steep temperature gradient, advection is causing $B$ to grow while $\alpha-e$ collisional and Nernst terms are preventing $B$ from penetrating further into the hot fuel, as can be seen from the bottom diagram of fig. \ref{f:3}. We describe the region where $\chi_{e}$ grows rapidly as self-insulating since the growth is driven by the dynamics of burn propagation. The top diagram of fig. \ref{f:3} also shows the final profiles of $\chi_{e}$ that are obtained if these two terms are neglected in the induction equation. Clearly, these terms play a crucial role in establishing the self-insulating region.

Finally, fig. \ref{f:4} shows profiles of quantities related to the $\alpha$ particles. It is found that $\chi_{\alpha}$ evolves in a similar way to $\chi_{e}$ with a spike in $\chi_{\alpha}$ occurring at the burn front. This causes a sharp drop in $F_{\bot}$ and a spike in $F_{\wedge} $ in the self-insulating region, while $\alpha$ heating propagates less far into the cold fuel compared with the unmagnetized case.

\newpage

\newpage

\subsection{Local Magnetization}\label{sec:3.3}
\begin{figure}[H]
\begin{center}%
\includegraphics*[width=1.0\columnwidth]{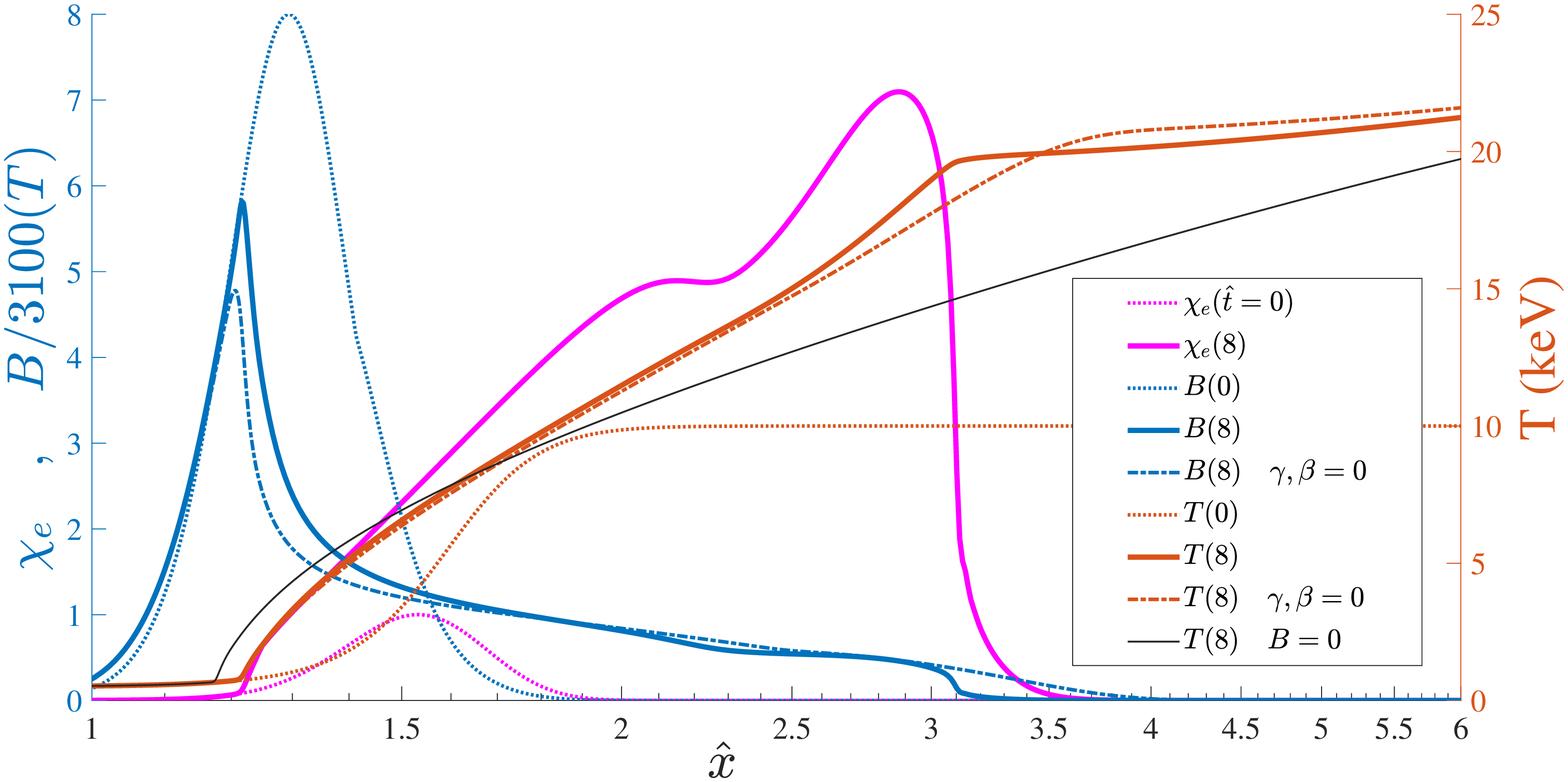}%
\hfil
\includegraphics*[width=1.0\columnwidth]{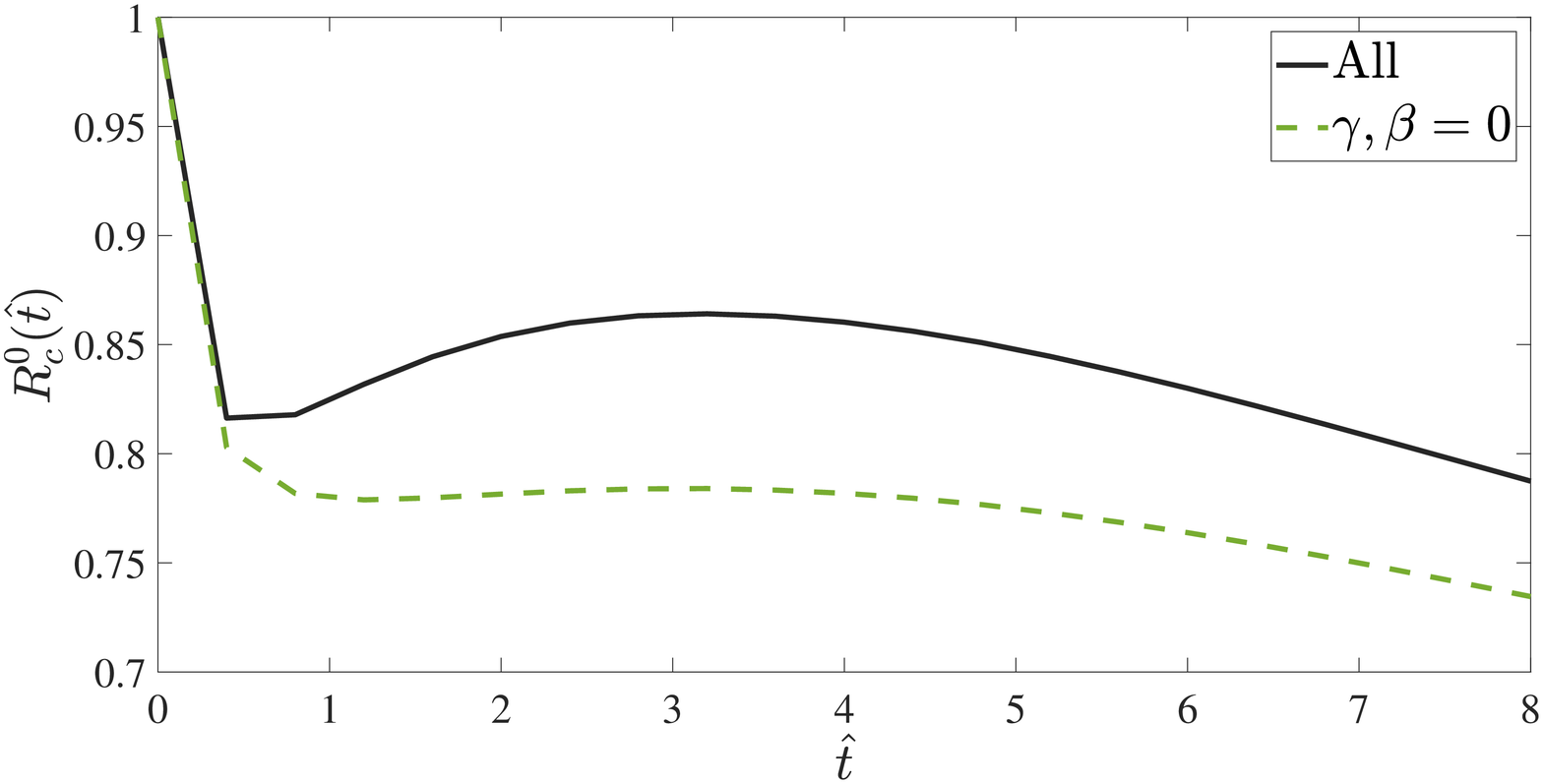}%
\vspace{-1em}
\end{center}
\caption[]{Top: Profiles of $\chi_{e}$, $B$ and $T$ at $\hat{t}=0$ and $\hat{t}=8\mathcal{T}_{h}$ for an initially local $B$ field profile. Also shown are $B$ and $T$ profiles for when Nernst and $\alpha-e$ collision terms are neglected in the induction equation ($\gamma,\beta = 0$) and the temperature profile for an unmagnetized plasma. Bottom: Ratio of cold fuel burn rate to unmagnetized rate as a function of time. Parameter values of $T_{c} = 500\,eV$, $T_{h} = 10\,keV$, $n_{h} = 10^{31}\,m^{-3}$ and $L_{T}=0.4=24\,\mu m$ were used.} \label{f:5}
\end{figure}

Integrated simulations of unmagentized ICF implosions have shown that self-generated $B$ fields can develop at the interface of hot and cold fuel.\cite{Walsh_PRL2017} In this section, we consider how such a local $B$ field could evolve during burn by assuming that the initial $\chi_{e}$ has a Gaussian profile that is narrower than $L_{T}$. An example of such a scenario is shown in fig. \ref{f:5} where the FWHM of the initial $\chi_{e}$ profile is $1/3L_{T}$ with a peak value of $\chi_{e}=1$.

The $B$ field is rapidly advected from cold to hot fuel by the propagating burn. However, the Nernst and $\alpha-e$ collision terms cause this advected $B$ field to maintain a steep front. This coincides with the front of the rapidly growing $\chi_{e}$ profile and a region of very steep temperature gradient. As in the case of initially uniform $\chi_{e}$, the growth in $\chi_{e}$ leads to the formation of a self-insulating layer between hot and cold fuel. This behaviour is rather counter-intuitive. Our usual expectation would be that the increasing temperature would increase the rate at which transport processes act to smooth out the steep gradients. Instead we see that the increasing temperature leads to increasing $\chi_{e}$ and a reduction in energy transport, such that the gradients become even steeper.

Figure \ref{f:5} also shows the reduction in burn rate of cold fuel as a function of time. We can see from this that even though the majority of fuel is unmagnetized the growth of $\chi_{e}$ during burn can significantly reduce the rate of burn propagation.

We note that the value of $\kappa_{\bot e}^{c}$, the electron thermal conductivity coefficient, is about an order of magnitude smaller at $\chi_{e}=7$ compared with $\chi_{e}=1$. Theoretical models of burn in unmagnetized ICF\cite{Guskov_1976,Hurricane_PPCF2018,Christopherson_POP2020} typically do not account for such rapid changes in conductivity but this may be necessary if the presence of self-generated $B$ fields is verified. Such fields may also alter the ablative stabilization of Rayleigh-Taylor instabilities during burn.\cite{Lobatchev_PRL2000}
\newpage

\section{Hall parameter growth rates}\label{sec:4}
The formation of a self-insulating layer with two steep temperature gradients during burn propagation is observed across a broad range of simulation parameters and initial $B$ field profiles. It is a result of the Hall parameter growing faster at the burn front than in the hot fuel. We can investigate its formation by considering the factors on which $\sigma_{\chi}$ depends. Starting with the definition $\chi_{e} = eB\tau_{ei}/m_{e}$ and assuming an isobaric, high-beta plasma with thermal pressure $P_{0} = 2nT$ we obtain
\begin{equation}
\sigma_{\chi}=\frac{D \ln\chi_{e}}{D \hat{t}} = \frac{1}{B}\frac{D B}{D \hat{t}}+\frac{g}{T}\frac{D T}{D \hat{t}},
\end{equation}
where $g = \left(5-3\left(\ln\Lambda_{ei}\right)^{-1}\right)/2\approx 5/2$. If we assume that $\chi_{e}$ is uniform, neglect bremsstrahlung losses ($\hat{q}_{\alpha}\gg\hat{P}$) and neglect resistive diffusion and Ettingshausen heat flow then using \eqref{e:3} and \eqref{e:4} results in
\begin{eqnarray}
\sigma_{\chi} = &&\frac{2\left(g-1\right)}{5P_{0}}\hat{q}_{\alpha}+
\hat{\beta}\left(\frac{\partial^{2} \ln T}{\partial \hat{x}^{2}}+\left(\frac{\partial \ln T}{\partial \hat{x}}\right)^{2}\right)+
\frac{2\left(g-1\right)}{5P_{0}}T\hat{\kappa}\left(\frac{\partial^{2} \ln T}{\partial \hat{x}^{2}}+\left(g+1\right)\left(\frac{\partial \ln T}{\partial \hat{x}}\right)^{2}\right)\nonumber\\
&&+\hat{\gamma}\left(\frac{\partial^{2} \ln \mathcal{E}_{\alpha}}{\partial \hat{x}^{2}}+\left(\frac{\partial \ln \mathcal{E}_{\alpha}}{\partial \hat{x}}\right)^{2}+\frac{\partial \ln T}{\partial \hat{x}}\frac{\partial \ln \mathcal{E}_{\alpha}}{\partial \hat{x}}\right),
\end{eqnarray}
The first term on the right-hand-side (the $\alpha$ heating term) will always be positive but the three remaining terms could be positive or negative, depending on the profiles of $T$ and $\mathcal{E}_{\alpha}$. This equation illustrates why the steep, opposing gradients in $\chi_{e}$ and $T$ are observed in figs. \ref{f:3} and \ref{f:5}: locations with $\partial^{2} \ln T/\partial \hat{x}^{2} <0$ lead to $\sigma_{\chi}<0$ and vice versa.

Assuming that the gradient length scales of $T$ and $\mathcal{E}_{\alpha}$ are similar, we can obtain the following estimates of the strengths of the different terms
\begin{eqnarray}
\frac{2\left(g-1\right)}{5P_{0}}\hat{q}_{\alpha} &\approx& \mathcal{T}_{H}6.29\times 10^{-5}\ln\Lambda_{\alpha e}\frac{\mathcal{E}_{\alpha}}{T_{keV}^{\frac{5}{2}}},\nonumber\\
\hat{\beta} &\approx& \mathcal{T}_{H}\frac{6.14\times 10^{12}}{\ln\Lambda_{ei}}\frac{T_{keV}^{\frac{7}{2}}}{L_{T\mu}^{2}P_{0G}}\frac{\beta_{\wedge}}{\chi_{e}},\nonumber\\
\frac{2\left(g-1\right)}{5P_{0}}T\hat{\kappa} &\approx& \mathcal{T}_{H}\frac{1.84\times 10^{12}}{\ln\Lambda_{ei}}\frac{T_{keV}^{\frac{7}{2}}}{L_{T\mu}^{2}P_{0G}}\left(\kappa_{\bot e}^{c}+ \sqrt{\frac{2m_{e}}{m_{i}}}\kappa_{\bot i}^{c}\right),\nonumber\\
\hat{\gamma} &\approx& \mathcal{T}_{H}\frac{4.09\times 10^{-2}}{\ln\Lambda_{\alpha e}}\mathcal{E}_{\alpha}\frac{T_{keV}^{\frac{7}{2}}}{L_{T\mu}^{2}P_{0G}^{2}}\Gamma_{\chi},\nonumber
\end{eqnarray}
where $T_{keV}$ is fuel temperature in $keV$, $P_{0G}$ is fuel pressure in $Gbar$ and $L_{T\mu}$ is temperature length scale in $\mu m$.

\begin{figure}[H]
\begin{center}%
\includegraphics*[width=1.0\columnwidth]{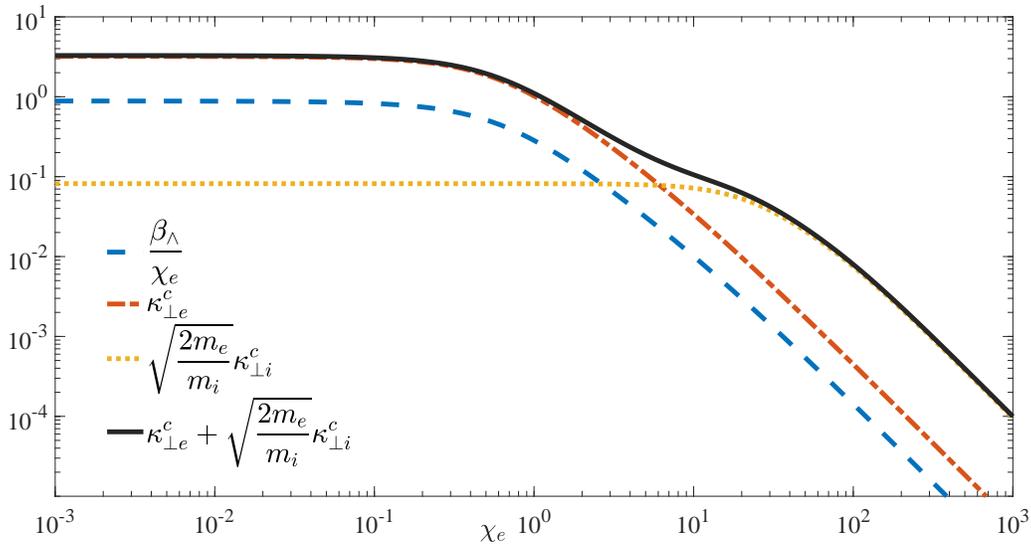}%
\vspace{-1em}
\end{center}
\caption[]{The variation of the dimensionless thermoelectricity and thermal conductivity coefficients with $\chi_{e}$. For $\chi_{e}> \sim6$, ion thermal conductivity is larger than electron thermal conductivity.} \label{f:10}
\end{figure}

From these relations we can see that the Nernst and thermal conductivity terms are very similar in magnitude. Their dependence on $\chi_{e}$ is shown in fig. \ref{f:10}. This figure illustrates that we expect the largest magnetization growth rates to occur at small values of $\chi_{e}$. It is likely that the cryogenic fuel layers of many MIF schemes will have initially low $\chi_{e}$ values due to the high density and so $\chi_{e}$ can increase rapidly in such regions as burn propagates.

\begin{figure}[H]
\begin{center}%
\includegraphics*[width=1.0\columnwidth]{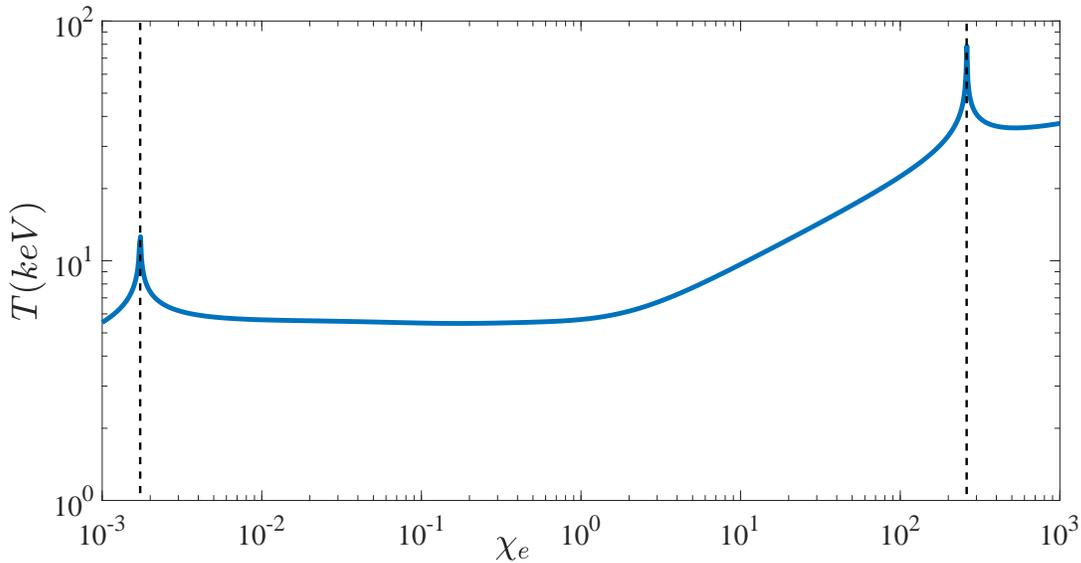}%
\vspace{-1em}
\end{center}
\caption[]{A plot of the minimum temperature for which the $\alpha$-e collisional term dominates the $\alpha$ heating term in the growth rate of $\chi_{e}$ for $L_{T\mu}=10$ and $P_{0G}=100$. This result may be scaled according to  \eqref{e:4.10b} for other values of $L_{T\mu}$ and $P_{0G}$. The dashed vertical lines indicate the values of $\chi_{e}$ for which $\Gamma_{\chi}=0$.} \label{f:10b}
\end{figure}

Finally, we can take the ratio of the $\alpha$ heating term to the $\alpha$-e collisional term to obtain a criterion (which is indepedent of $\mathcal{E}_{\alpha}$) for when the latter term is larger
\begin{equation}
T_{keV}^{6}\left|\Gamma_{\chi}\right|>\frac{\left(\L_{T\mu}P_{0G}\ln\Lambda_{\alpha e}\right)^{2}}{650}.\label{e:4.10b}
\end{equation}
From this formula we can, for given values of $\chi_{e}$, $L_{T\mu}$ and $P_{0G}$, calculate the temperature above which the $\alpha$-e collisional term is larger. An example of this is shown in fig. \ref{f:10b}.


\section{Conclusions}\label{sec:5}
In this work we have studied the propagation of a subsonic thermonuclear burn wave across a magnetic field in a high beta plasma using an MHD model. We can summarize our findings as follows:
\begin{itemize}
\item[(i)] Ablation of cold fuel due to thermonuclear burn advects $B$ field from cold to hot fuel at the burn front, while the Nernst and $\alpha-e$ collision terms transport $B$ field from hot to cold fuel. These combined transport effects confine $B$ field in the burn front region and cause both compression and rarefaction of $B$.
\item[(ii)] The electron and $\alpha$ Hall parameters, $\chi_{e}$ and $\chi_{\alpha}$ , can grow rapidly at the burn front due to $\alpha$ heating, magnetic field transport and thermal conduction processes. The growth rate at the burn front can be significantly larger than in the hot fuel, causing a self-insulating layer to form at the burn front.
\item[(iv)] The temperature profile of a magnetized burn front is steeper than the unmagnetized case due to the suppression of thermal conduction. Furthermore, the formation of a self-insulating layer can lead to very steep temperature gradients in localized regions of the burn front.
\item[(iv)] A magnetized burn front reduces the rate at which cold fuel is burnt due to the self-insulating effect.
\end{itemize}
These results demonstrate that careful modelling of the $B$ field transport during burn is required to accurately simulate the propagating burn wave and predict the energy gain achieved. However, some of the assumptions in our idealized model prevent us from making quantitative predictions for how $B$ field transport could affect specific MIF schemes. The isobaric condition restricts us to regimes of subsonic burn (deflagration). Our results prompt the question of whether the formation of a self-insulating layer at the burn front could cause a transition to detonation of the hot fuel. Furthermore, the planar geometry used here is suitable for idealized studies of burn front dynamics since the hot fuel at infinity is unaffected by the cold fuel. It is likely that there will be a more complex interplay between hot fuel and burn front in finite sized targets. Questions such as these will be addressed in future work using integrated simulations.

The results also provide further evidence\cite{Slutz_PRL2012,Velikovich_ICOPS2012} that an optimum magnetic field strength and profile shape could exist for a given MIF scheme. Such a field would need to provide the correct amount of magnothermoinsulation during compression and ignition, followed by sufficient confinement of $\alpha$ particles during burn propagation without unnecessary slowing of the burn wave. The strength and shape of this optimum magnetic field and how it can be generated remain open questions.

\nocite{*}
\bibliographystyle{unsrt}



\end{document}